\documentclass[nofootinbib,preprint,superscriptaddress,showpacs,showkeys,preprintnumbers,amsmath,amssymb,aps,pre]{revtex4-2}

\usepackage[utf8]{inputenc} 
\usepackage{footnote}
\usepackage{epsfig}

\usepackage{natbib}
\usepackage{booktabs}
\usepackage{amsmath}
\usepackage{times}
\usepackage{epstopdf} 
\usepackage{graphicx}
\usepackage{color}

\usepackage{afterpage}

\begin{document}

\title{A simple method to enlarge a basin of attraction using a memristive function}

\author{Alexandre R. Nieto}
\email[]{alexandre.rodriguez@urjc.es}
\affiliation{Nonlinear Dynamics, Chaos and Complex Systems Group, Departamento de
F\'{i}sica, Universidad Rey Juan Carlos, Tulip\'{a}n s/n, 28933 M\'{o}stoles, Madrid, Spain}

\author{Rubén Capeáns}
\affiliation{Nonlinear Dynamics, Chaos and Complex Systems Group, Departamento de
	F\'{i}sica, Universidad Rey Juan Carlos, Tulip\'{a}n s/n, 28933 M\'{o}stoles, Madrid, Spain}

\author{Miguel A.F. Sanju\'{a}n}
\affiliation{Nonlinear Dynamics, Chaos and Complex Systems Group, Departamento de
F\'{i}sica, Universidad Rey Juan Carlos, Tulip\'{a}n s/n, 28933 M\'{o}stoles, Madrid, Spain}

\date{\today}

\begin{abstract}
	This study presents an innovative approach to chaotic attractor stabilization introducing a memristor in discrete dynamical systems. Using the Hénon map as a test case, we replace a system parameter with a memristive function governed by a sigmoid activation function. The method relies on leveraging attractors with larger basins of attraction to attract the orbits and guide them towards the desired chaotic attractor. The effectiveness of the method is confirmed through numerical simulations, showing substantial enhancement in attractor stability without requiring explicit parameter control.
\end{abstract}



\maketitle



\newpage

\section{Introduction} \label{sec1}
In $1971$, Prof. Chua theoretically predicted the existence of the memristor \cite{Chua71}—a resistor whose resistance depends on the amount of electric charge that has flowed through it. Almost $40$ years later, in $2008$, Hewlett-Packard Labs demonstrated a nanoscale titanium dioxide memristor \cite{Strukov08}.

In addition to their numerous technological applications in the field of nanoelectronics \cite{Pickett09,Brown23}, memristors have been extensively studied in the context of chaos theory. Beyond the many studies exploring its effects on systems of nonlinear ODEs, the impact of memristors on discrete chaotic systems has recently attracted considerable attention \cite{Wang22,Rong22,Peng20}. Some of these recent works have used paradigmatic maps, such as the Hénon map and the Lozi map, to demonstrate that the incorporation of a memristor can completely alter the system's dynamics.

A continuous model of a memristor can be defined as
\begin{align}	
	\begin{aligned}
		&v(t)=M(q(t))i(t), \\
		&\dot{q}=i(t),
	\label{Memristor}
	\end{aligned}
\end{align} 
where $i(t)$ is the current, $q(t)$ the charge, $v(t)$ the voltage, and $M(q(t))$ the memristance. By discretizing the derivatives, a discrete memristor model can be expressed as
\begin{align}
	\begin{aligned}		
	&	v_{n}=M(q_n)i_n, \\
	&	q_{n+1}=q_n+hi_n,
	\end{aligned}		
\label{Memristor2}
\end{align} 
where $h$ is the time step and the memristive function $M(q_n)$ is often modeled as a sigmoid function.

Building on this simple discrete memristor model, we present a method to enlarge the basin of attraction of a chaotic attractor by replacing a system parameter with a memristive function. For all numerical simulations, we have used the Hénon map as the model, demonstrating the method by enlarging the basin of attraction of the Hénon attractor.

The structure of this paper is as follows. In Sec.~\ref{sec2} we explain the method. Numerical simulations using the Hénon map, which demonstrate how the basins of attraction are effectively enlarged, are presented in Sec.~\ref{sec3}. Finally, in Sec.~\ref{sec4}, we summarize the key findings.

\section{The method}\label{sec2}
For this research we have used the Hénon map as a model, so we will use the same system to explain the method. The Hénon map is a paradigmatic two-dimensional map given by
\begin{align}
{\cal{H}}(a,b,c):	\begin{cases}		
		x_{n+1}=1+ax^2+cy_n, \\
		y_{n+1}= bx_n,\\
	\end{cases}		\label{henon}
\end{align} 
where $a,b,c\in\mathbb{R}$ are parameters. In the canonical Hénon map $c=1$, but we have included a free parameter for the purposes of this research.

\begin{figure}[h!]
	\centering
	\includegraphics[clip,height=7.2cm,trim=0cm 0cm 0cm 0cm]{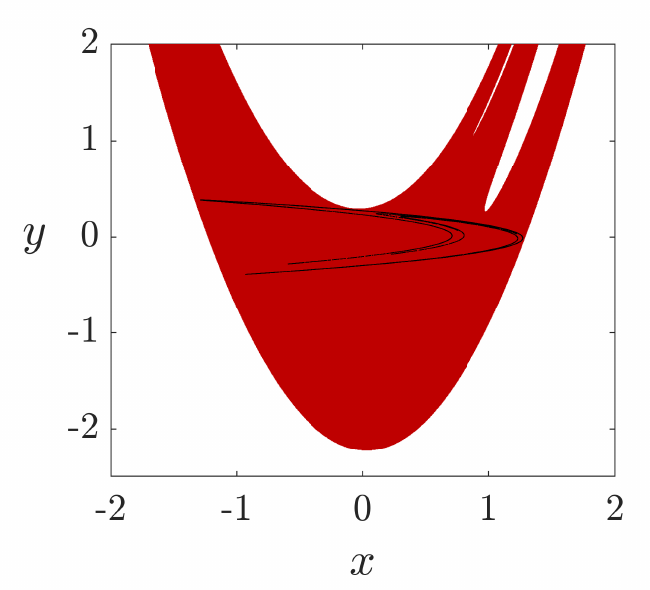}
	\caption{In black, chaotic attractor $\cal{C}$ of the Hénon map given by Eq.~(\ref{henon}) with parameters $a=1.4$, $b=0.3$, and $c=1$. In red, basin of attraction of $\cal{C}$. White color represents initial conditions that generate unbounded orbits that diverge to $\pm\infty$.   }
	\label{Fig1}
\end{figure}

For parameter values $a=1.4$, $b=0.3$, and $c=1$, the only attractive set of $\cal{H}$ is a chaotic attractor $\cal{C}$. Thus, if we define an arbitrary phase space region $\Omega\in[-10,10]\times[-50,50]$, all initial conditions outside the basin of attraction of $\cal{C}$ generate unbounded orbits, meaning that they leave $\Omega$. We show the chaotic attractor and its basin of attraction in Fig.~\ref{Fig1}. The area of this basin in $\Omega$ is $\mu[{\cal{B}}({\cal{C}})]=21.8$. Numerically, the area of the basin is calculated as the fraction of initial conditions belonging to the basin, multiplied by the total area of $\Omega$. 

The objective of our method is to enlarge the basin of attraction shown in in Fig.~\ref{Fig1}. If we achieve this, a wider range of initial conditions will generate orbits that converge to $\cal{C}$, so the stability of the chaotic attractor would be enhanced. 

\newpage
For this purpose, we replace one of the parameters of $\cal{H}$ by a memristive function $M(z_n)$, where $z$ is the inner state of the memristor (charge in Eq.~(\ref{Memristor2})). To maintain simplicity, we modify only one parameter at a time, producing three variants of the memristive Hénon map given by

\begin{align}
{\cal{H}}_a(M,b,c):	\begin{cases}		
		x_{n+1}=1+M(z_n)x^2+cy_n, \\
		y_{n+1}= b x_n,\\
		z_{n+1}=z_n+x_n.
	\end{cases}		\label{memA}
\end{align} 
\begin{align}
	{\cal{H}}_b(a,M,c):	\begin{cases}		
		x_{n+1}=1+ax^2+cy_n, \\
		y_{n+1}= M(z_n) x_n,\\
		z_{n+1}=z_n+x_n.
	\end{cases}		\label{memB}
\end{align} 
\begin{align}
	{\cal{H}}_c(a,b,M):	\begin{cases}		
		x_{n+1}=1+ax^2+M(z_n)y_n, \\
		y_{n+1}= b x_n,\\
		z_{n+1}=z_n+y_n.
	\end{cases}		\label{memC}
\end{align} 
In Eqs.~(\ref{memA}-\ref{memC}), we define the memristive function as
\begin{equation}
	M(z_n)=p_u+(p_t-p_u)\varphi(z_n), \label{memform}
\end{equation} 
 where $p\in\{a,b,c\}$, and the subscripts $t$ and $u$ stand for `target' and `undesired', respectively, indicating that the method aims for $M(z_n)$ to converge to the target parameter $p_t$ rather than the undesired parameter $p_u$. The activation function $\varphi(z_n)$ is the sigmoid
\begin{equation}
	\varphi(z_n) =\frac{1}{1+e^{-z_n/\tau}}, \label{logistic}
\end{equation} 
where $\tau$ determines the steepness of the curve. 

\begin{figure}[h!]
	\centering
	\includegraphics[clip,height=5.6cm,trim=0cm 0cm 0cm 0cm]{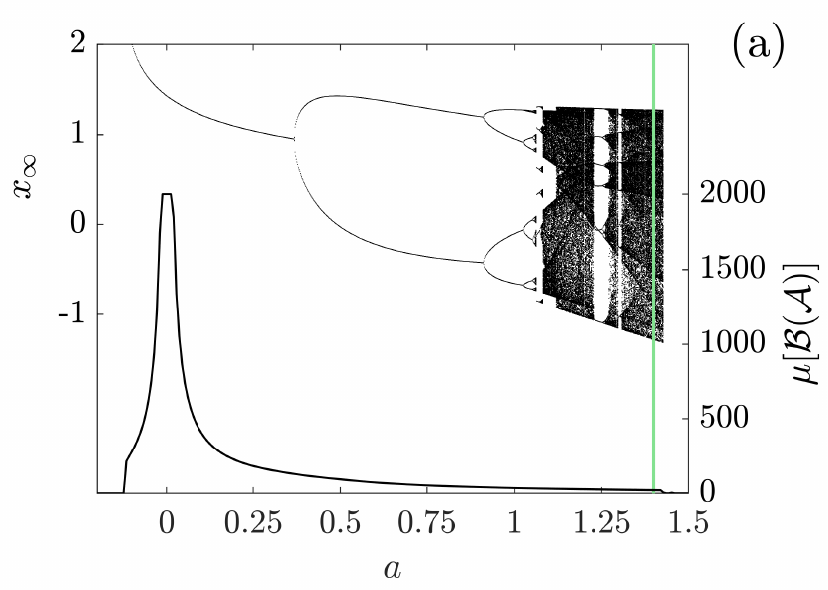}
	\includegraphics[clip,height=5.6cm,trim=0cm 0cm 0cm 0cm]{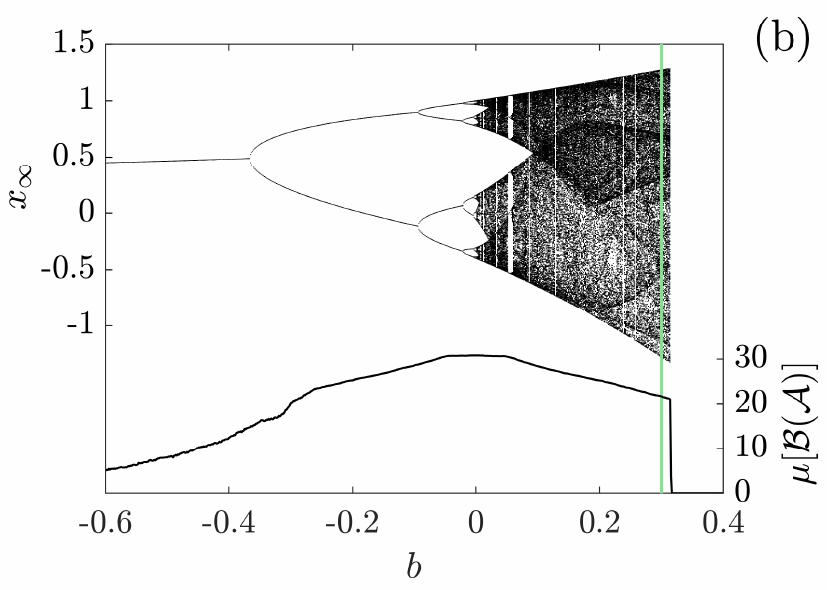}
	\includegraphics[clip,height=5.6cm,trim=0cm 0cm 0cm 0cm]{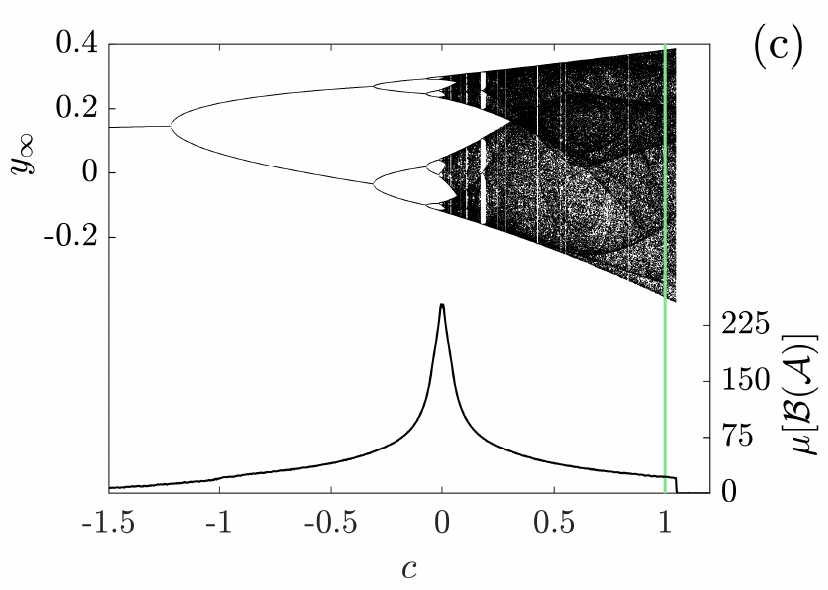}
	\caption{Bifurcation diagrams illustrating the asymptotic behavior of the H\'enon map as a function of the system parameters. (a) Bifurcation diagram with respect to parameter $a$, (b) with respect to parameter $b$, and (c) with respect to parameter $c$. The green vertical line in each plot indicates the position of the target chaotic attractor $\mathcal{C}$. Below each bifurcation diagram, the area of the basin of attraction, $\mu[{\cal{B}}({\cal{A}})]$, is shown as a function of the corresponding parameter $p \in \{a,b,c\}$. This lower curve represents how the size of the basin of attraction changes with the parameter value, showing characteristic peaks that indicate regions where the attractors have larger basins of attraction.}
	\label{Fig2} 
\end{figure}

Since $\varphi(z_n) \in (0,1)$, the structure of the memristive function given by Eq.~(7) guarantees $M(z_n) \in (p_u,p_t)$. When $z_n$ diverges to $+\infty$, $M(z_n)$ converges to the target parameter $p_t$ ($a_t = 1.4$, $b_t = 0.3$, or $c_t = 1$ in our case), leading to orbit stabilization in the target attractor $\mathcal{A}(p_t) = \mathcal{C}$. Conversely, if $z_n$ diverges to $-\infty$, $M(z_n)$ converges to an undesired parameter $p_u$, resulting in orbit stabilization in the attractor $\mathcal{A}(p_u)$. The final convergence to either attractor depends on the accumulated value of $x$ or $y$ along the orbit. Therefore, the asymptotic behavior of the memristive map ${\cal{H}}_p$ can be understood by simply observing the appropriate coordinate in a bifurcation diagram that represents $\cal{A}$ as a function of $a$, $b$, or $c$. These diagrams are shown in Fig.~\ref{Fig2}. The left $y$-axis represents the asymptotic value of the coordinate, while the right $y$-axis represents the area of the basin of attraction of the attractor ${\cal{A}}(p)$ for each parameter value. The vertical green line indicates the position of the target attractor $\cal{C}$.

For all three parameters, there exist attractors $\mathcal{A}(p)$ with basins of attraction larger than that of $\mathcal{C}$. Our method exploits these larger basins through the following mechanism:

\begin{enumerate}
	\item The initial value of the memristive function $M(z)$ is the midpoint of the parameter range
	\[M(0) = \frac{p_t + p_u}{2} \equiv p_0.\]
	
	\item If $p_m$ is the parameter for which the basin of attraction of ${\cal{A}}(p)$ is the largest (see absolute maxima in Fig.~\ref{Fig2}), our best choice for $p_u$ is the one that satisfies $p_0=p_m$. Since $\mu[\mathcal{B}(\mathcal{A}(p_0))] > \mu[\mathcal{B}(\mathcal{C})]$, by doing so we ensure that more initial conditions are captured. Note that we treat $p_u$ as a free parameter since $p_t$ is fixed once we choose a target attractor.
	
	\item To achieve $p_0 = p_m$, we can derive the optimal value of $p_u$ (denoted as $p_u^o$):
	\[p_m = \frac{p_t + p_u^o}{2} \implies p_u^o = 2p_m - p_t.\]
\end{enumerate}

This choice of optimal $p_u$ ensures maximal initial capture of orbits. However, for the method to be successful, the memristive function $M(z_n)$ must converge to $p_t$ rather than $p_u$. If $M(z_n)$ were to converge to $p_u$, the basin of attraction would vanish ($\mu[\mathcal{B}(\mathcal{C})] = 0$), and our attempt to enhance the stability of $\mathcal{C}$ would have been unsuccessful.

In the Hénon map, the average values of coordinates $x$ and $y$ across all attractors (including $\mathcal{C}$) are positive. This property ensures that $M(z_n)$ converges to $p_t$, making $\mathcal{C}$ the only asymptotically stable attractor of the map $\mathcal{H}_p$. For systems where coordinate values might not be positive, the form of $M(z_n)$ in Eq.~(\ref{memform}) would need to be adjusted accordingly to ensure convergence to the desired attractor.

\section{Numerical simulations in the memristive Hénon maps}\label{sec3}
This section presents the results of numerical simulations performed using the memristive Hénon maps given by Eqs.~(\ref{memA}-\ref{memC}). In each case, we have computed the area of the basin of attraction of $\cal{C}$ for different values of $p_u$. Without loss of generality, we set $\tau = 100$ in Eq.~(\ref{logistic}). This choice ensures sufficiently slow parameter variation, preventing physically unrealistic discontinuous jumps. Our numerical analysis confirms that the qualitative behavior remains consistent across a wide range of $\tau$ values, from $\tau = 0.1$ to $\tau = 1000$. While $\tau$ influences the transient time required for convergence to $\mathcal{C}$, its effect on the area of the basins of attraction is negligible, at least in the context of the H\'enon map.

\begin{figure}[h!]
	\centering
	\includegraphics[clip,height=6.8cm,trim=0cm 0cm 0cm 0cm]{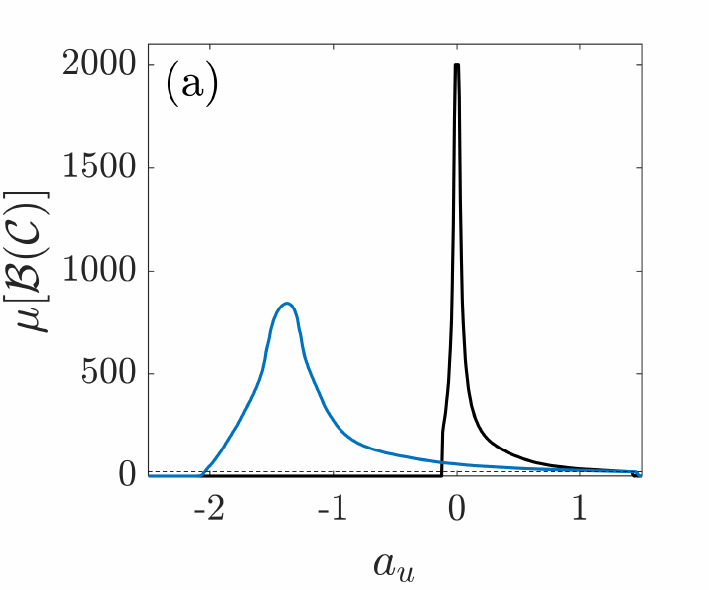}
	\includegraphics[clip,height=6.8cm,trim=0cm 0cm 0cm 0cm]{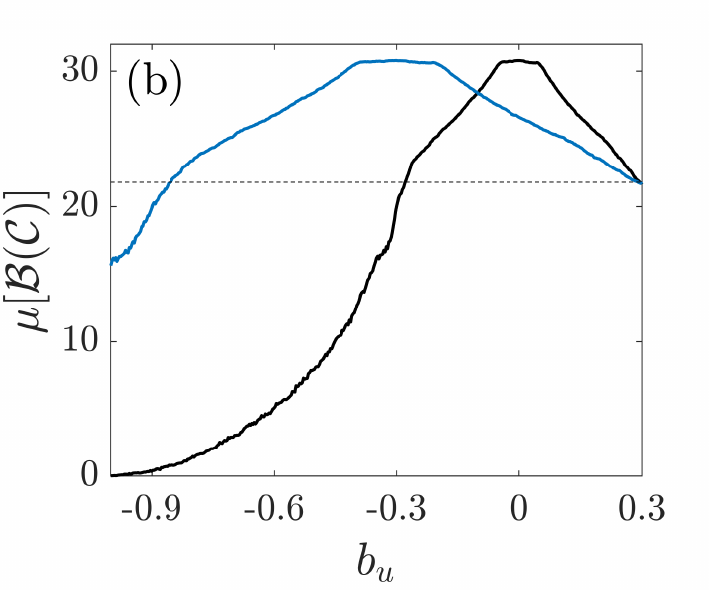}
	\includegraphics[clip,height=6.8cm,trim=0cm 0cm 0cm 0cm]{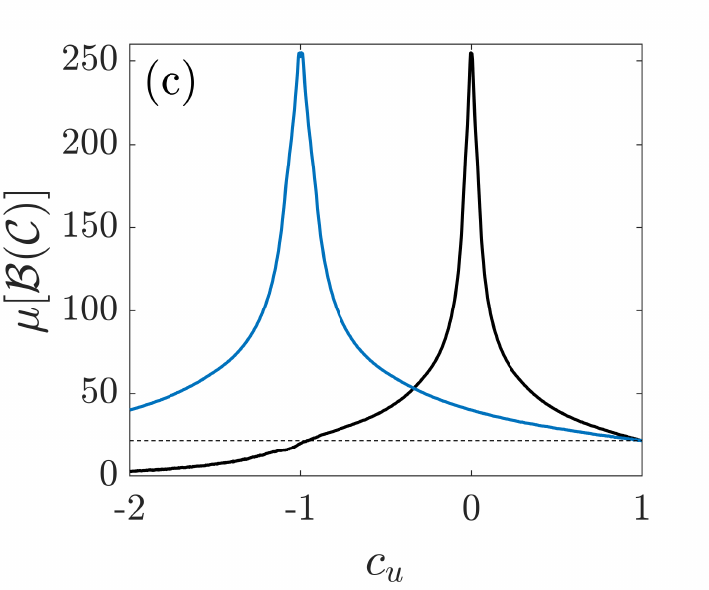}
	\caption{The blue curves represent the area of the basins of attraction of $\cal{C}$ as a function of the free parameter $p_u$ in the memristive map given by (a) Eq.~(\ref{memA}), (b) Eq.~(\ref{memB}), and (c) Eq.~(\ref{memC}). The black curve represents the area of the basins of attraction of the different attractors $\cal{A}$ in the Hénon map given by Eq.~(\ref{henon}) as a function of the parameters (a) $a$, (b) $b$, and (c) $c$. The horizontal dashed line indicates the area of the basin of attraction of the chaotic attractor $\cal{C}$ in the Hénon map (i.e., the map without memristor). Every value of $\mu[\cal{B}(C)]$ over this line implies that the basin is enlarged.  }
	\label{Fig3} 
\end{figure}

The results for the area of the basin of attraction of $\mathcal{C}$ are shown with blue curves in Fig.~\ref{Fig3}, where each panel (a), (b), and (c) corresponds to the numerical simulation performed by replacing one of the three parameters $a$, $b$, and $c$ with the memristive function $M(z_n)$. The black curves in each panel represent the area of the basins of the attractors $\mathcal{A}$ in the H\'enon map as a function of the corresponding parameter, which are identical to those shown at the bottom of the panels in Fig.~\ref{Fig2}. By displaying both curves together, a strong correlation between them becomes evident. The horizontal dashed line marks the area of the basin of $\mathcal{C}$ in the H\'enon map, providing a reference value for comparison.

\newpage
The data shown in Fig.~\ref{Fig3} demonstrates that the method successfully enlarges the basin of attraction of $\cal{C}$. Furthermore, as reasoned in the previous section, the relationship $p_{u}^o=2p_m-p_t$ holds. For instance, in the case shown in Fig.~\ref{Fig3}(c), the parameters are $a_t=1$ and $a_m=0$, so the expected optimal value would be $p_{u}^o=-1$, as confirmed by the numerical simulation.

The effectiveness of the method depends entirely on the area of the basins of attraction of the attractors in the Hénon map. In the best-case scenario, as seen in Figs.~\ref{Fig3}(b-c), we achieve $\mu[{\cal{B}}{(\cal{C})}]=\mu[{\cal{B}}{(\cal{A}}(p_m))]$. That is, in the optimal case, all initial conditions attracted by ${\cal{A}}(p_m)$ eventually reach $\cal{C}$. However, this is not always the case, as demonstrated in Fig.~\ref{Fig3}(a). The maximum values of the area $\mu[{\cal{B}}{(\cal{C})}]$ shown in the different panels of Fig.~\ref{Fig3} are $844.3$, $30.8$, and $254.8$, respectively. Therefore, depending on which parameter of the map $\cal{H}$ is replaced with the memristor, $\mu[{\cal{B}}({\cal{C}})]$ can increase by a factor $38.7$ (parameter $a$), $1.4$ (parameter $b$), or $11.7$ (parameter $c$).  

\begin{figure}[h!]
	\centering
	\includegraphics[clip,height=6.8cm,trim=0cm 0cm 0cm 0cm]{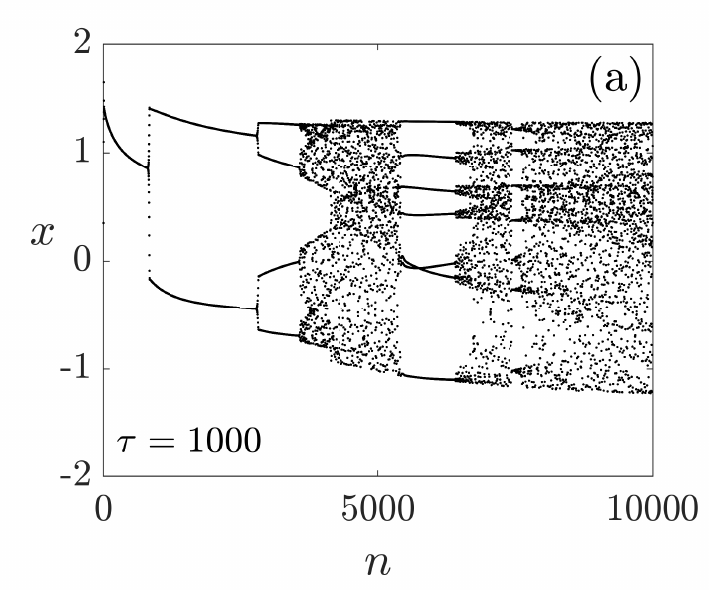}
	\includegraphics[clip,height=6.8cm,trim=0cm 0cm 0cm 0cm]{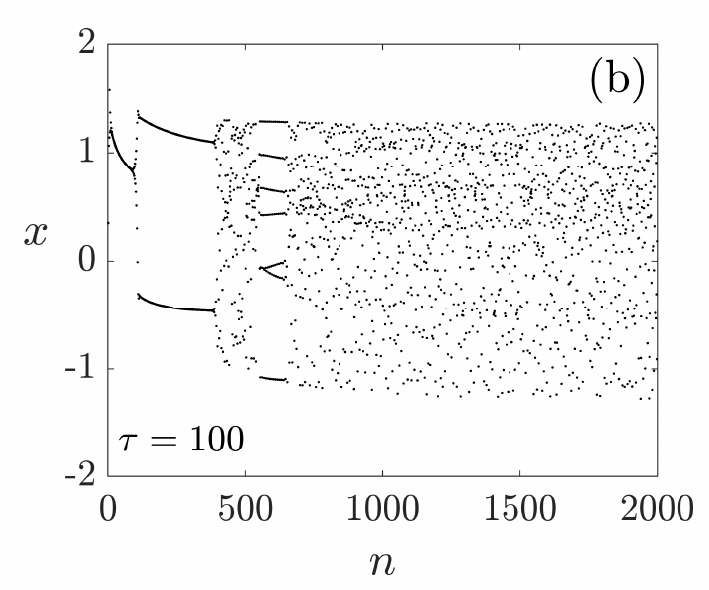}
	\caption{Transition of the orbit from the attractor ${\cal{A}}(p_m)$ (a fixed point) to the target chaotic attractor $\cal{C}$ using (a) $\tau=1000$ and (b) $\tau=100$ in the activation function. The memristive Hénon map is defined by Eq.~(\ref{memA}) with $b=0.3$, $c=1$, $a_t=1.4$, and $a_u=-1.4$. The initial condition is $(x_0,y_0)=(0.35,3)$.}
	\label{Fig4} 
\end{figure}

We conclude this work by illustrating the transition of the orbit from the attractor ${\cal{A}}(p_m)$ to the chaotic attractor $\cal{C}$, using two different values of the parameter $\tau$ in the activation function of $M(z)$. These orbits are shown in Fig.~\ref{Fig4}, where the memristive Hénon map given by Eq.~(\ref{memA}) with $b=0.3$, $c=1$, $a_t=1.4$, and $a_u=a_u^o=-1.4$ is considered. In both simulations we have set $(x_0,y_0)=(0.35,3)$. This initial condition generates an unbounded orbit in the Hénon map, but successfully reaches $\cal{C}$ in the memristive map, regardless of the value of $\tau$. 

\section{Conclusions and discussion}\label{sec4}
In conclusion, in this work we have shown a simple method to enhance the stability of a chaotic attractor using a memristor. The method involves replacing a parameter of the system with a memristive function. This modification enables transient dynamics that facilitates the attraction of orbits towards attractors with large basins of attraction, ultimately converging to the desired attractor.

We have illustrated the method using the Hénon map, but the features of the system that enable the method to work are quite general in chaotic systems. The key requirement is the existence of parameter values where the system exhibits attractors with basins of attraction larger than the target basin. Implementation in other systems requires only adjusting the form of the memristive function $M(z_n)$ according to the bifurcation direction and the signs of the attractor coordinates.

We are aware that the goal of this method could be achieved similarly using a parameter drift. However, we believe that the idea of using a memristor is noteworthy because it achieves the goal without requiring absolute control over the system's parameters. Parameter drift for control purposes may imply that the parameter value can be set at will, whereas in the case of the memristor, it is the system's own dynamics that stabilizes the attractor.

 \section*{Acknowledgments}
This work has been financially supported by MCIN/AEI/10.13039/501100011033 and by “ERDF A way of making Europe” (Grant No. PID2023-148160NB-I00).

\section*{Disclosure statement}
No potential conflict of interest was reported by the author(s).

\section*{Data availability statement}
The data that support the findings of this study are available from the corresponding author, ARN, upon reasonable request.

\end{document}